\pdfoutput=1

\documentclass[11pt]{article}

\usepackage[preprint]{acl}

\usepackage{times}
\usepackage{latexsym}

\usepackage[T1]{fontenc}

\usepackage[utf8]{inputenc}

\usepackage{microtype}

\usepackage{inconsolata}

\usepackage{graphicx}
\usepackage{amsmath}
\usepackage{float}
\usepackage{enumitem}
\usepackage{multirow}

\usepackage[utf8]{inputenc} 
\usepackage[T1]{fontenc}    
\usepackage{hyperref}       
\usepackage{url}            
\usepackage{booktabs}       
\usepackage{amsfonts}       
\usepackage{nicefrac}       
\usepackage{microtype}      
\usepackage{xcolor}         
\usepackage{graphicx}
\usepackage{wrapfig}
\usepackage{amsmath}
\usepackage{amssymb}
\usepackage{mathtools}
\usepackage{amsthm}
\usepackage{multirow}
\usepackage{makecell}
\usepackage{caption}
\usepackage{subcaption}
\usepackage{wrapfig}
\usepackage{graphicx}

\usepackage{pifont}
 \title{Seq1F1B: Efficient Sequence-Level Pipeline Parallelism for \\Large Language Model Training}

\begin{document}
\def\thu{NLP Group, DCST, IAI, BNRIST, Tsinghua University, Beijing, China.}
\def\bupt{Beijing University of Posts and Telecommunications, Beijing, China.}
\author{
  Ao Sun${^{1,*}}$ \hspace{4mm} Weilin Zhao$^{2, *}$\hspace{5mm} 
  Xu Han${^{2,\dag}}$\hspace{5mm} 
  \AND
  Cheng Yang${^{1, \dag}}$\hspace{4mm}
  Xinrong Zhang ${^{2}}$\hspace{4mm}
   Zhiyuan Liu${^{2}}$ \hspace{4mm} Chuan Shi$^{1}$ \hspace{4mm} Maosong Sun$^{2}$ \\
  $^1$ \bupt \hspace{3pt}
  \\
  $^2$ \thu \hspace{3pt}
  \\
  \texttt{\{maydomine, yangcheng\}@bupt.edu.cn} \\
  \texttt{\{zwl23,zxr19\}@mails.tsinghua.edu.cn} \\
  \texttt{han-xu@tsinghua.edu.cn}
  \\
}

\maketitle
\footnotetext[1]{$^*$ Indicates equal contribution.}
\footnotetext[2]{$\dag$ Indicates corresponding author.}
\begin{abstract}

Training large language models (LLMs) heavily relies on distributed training strategies, among which pipeline parallelism (PP) plays a crucial role. As training sequences extend to 32K or even 128K tokens, current PP methods face severe bottlenecks, including substantial pipeline bubbles and high memory footprint, greatly hindering training throughput and model scalability. 
This paper introduces a sequence-level one-forward-one-backward (1F1B) PP method, named Seq1F1B, tailored for training LLMs on long sequences with high training throughput and memory efficiency. 
Unlike typical PP methods, which adopt batch-level pipeline schedule, Seq1F1B schedules the pipeline of training LLMs at the sequence level. It uses a computational strategy to partition sequences appropriately, significantly reducing pipeline bubbles and memory footprint. 
Compared to competitive PP baselines such as Megatron 1F1B PP, Seq1F1B achieves 1.14$\times$ training throughput with half memory footprint compared to competitive baseline.
Notably, Seq1F1B trains an LLM with 30B parameters on sequences up to 64K tokens using 64$\times$NVIDIA A100 GPUs without using recomputation strategies, a feat unachievable with existing methods.
We will release our code to facilitate further research and development in LLM training on long sequences. 
\end{abstract}

\section{Introduction}
\label{intro}
Efficient distributed strategies ~\cite{shoeybi2019megatron,li2020pytorch,megatron2021} play a crucial role in training large language models (LLMs), and these LLMs have revolutionized various NLP tasks in recent years~\cite{touvron2023llama,reid2024gemini,jiang2024mixtral,anil2023palm}. 
Among these strategies, pipeline parallelism (PP)~\cite{gpipe,megatron2021} stands out due to its low communication bandwidth requirement and great computing resource scalability, and it can be easily integrated with other strategies such as data parallelism (DP)~\cite{li2020pytorch,rasley2020deepspeed} and tensor parallelism (TP)~\cite{shoeybi2019megatron,megatron-v3}. 


PP involves partitioning a model into multiple stages, with each computing device processing a stage consisting of consecutive layers.
This paradigm inherently leads to ``bubbles''---the idle time caused by the execution topology between the sharded layers. 
Several ingenious pipeline schedule strategies have been proposed to address this bubble problem.
GPipe~\citep{gpipe} reduces bubbles by splitting each batch of training sequences into micro-batches, coming at the cost of increased memory usage, as each pipeline stage must store the intermediate states of all micro-batches generated during forward passes until backward passes are completed. 
To address the high memory demand of GPipe, one-forward-one-backward (1F1B) methods are proposed~\citep{pipedream,fan2021dapple,megatron2021}. 1F1B methods make backward passes have higher execution priority than forward passes and schedule backward passes in advance without affecting final results. Owing to this, the memory demand for storing intermediate states can be reduced without adding extra bubbles. 
Generally, optimizing PP relies on handling the trade-off between bubble ratio and memory footprint. 

Recently, some efforts~\cite {long,reid2024gemini} have noticed that long-sequence training benefits LLMs in many aspects. However, it takes work to train LLMs on long sequences due to the quadratic time and memory complexities of Transformer attention modules in terms of sequence~\cite {vaswani2017attention}. In distributed training scenarios, long sequences also cause various parallelism methods to fail. For GPipe and existing 1F1B methods, whose minimal schedulable unit is micro-batch, inevitably face the memory overflow caused by just a single micro-batch, as training sequences extend to extremely long lengths. Long sequences make balancing bubble ratio and memory footprint more challenging for PP methods.


In this paper, we introduce a Sequence-Level 1F1B (Seq1F1B) PP method. This method capitalizes on the causal self-attention mechanism of LLMs to schedule pipeline stages at the sequence level. In contrast to existing 1F1B methods~\cite{megatron2021,qi2024zero-bubble}, Seq1F1B offers significant efficiency and memory benefits. 
Specifically, splitting sequences into sub-sequences allows for a significant reduction in memory footprint since only the intermediate states of sub-sequences rather than micro-batches need to be retained. Scheduling the pipeline at the sequence level yields more stages and thus reduces the bubble ratio.
While, the causal nature of LLMs also causes a dependency between the forward and backward passes of different sub-sequences, i.e., the forward passes of later sub-sequences rely on earlier ones, and vice versa for the backward passes of early sub-sequences, bringing the challenge for the pipeline schedule. To this end, we introduce a partially ordered queue in Seq1F1B to replace the first-in-first-out (FIFO) queue used in existing 1F1B methods and reorganize the pipeline schedule, so that we can preserve the exact execution dependencies between forward and backward passes while providing synchronous parallelism. 
To further improve Seq1F1B, we propose a strategy for balancing the workload across sub-sequences rather than simply splitting sequences evenly along the sequence dimension.

Sufficient experiments demonstrate that Seq1F1B significantly outperforms recent popular 1F1B methods~\cite{megatron2021,fan2021dapple} in terms of memory efficiency and training throughput for training LLMs, with the sequence length ranging from 16K to 128K and the model size ranging from 2.7B to 32B. 
As the sequence length increases, the efficiency of Seq1F1B becomes more pronounced. Seq1F1B supports efficiently training an LLM with 30B parameters on sequences up to 64K tokens using 64$\times$NVIDIA A100 GPUs without using recomputation strategies, which is unachievable with existing PP methods.

\section{Related Work} 

Training LLMs requires using a mixture of parallelism strategies, the most important of which are DP, TP, and PP. 
PP focuses on partitioning a model into multiple stages, with each device processing a stage consisting of consecutive layers, which is the core strategy to scale more devices to train LLMs. For PP, pipeline schedules can be broadly categorized into two main types: synchronous schedules and asynchronous schedules. Asynchronous schedules such as asynchronous PipeDream~\citep{pipedream} and PipeMare~\citep{yang2021pipemare}
can achieve bubble-free results but suffer from the performance degradation of final trained models because they use outdated parameters to compute gradient updates. 
As for synchronous schedules, GPipe~\citep{gpipe,li2021terapipe} and 1F1B~\citep{fan2021dapple,megatron2021,pipedream-flush} are the most commonly used pipeline schedules following synchronous settings. They achieve much fewer bubbles as the number of micro-batch increases and guarantee mathematical equivalent to the original training process.
Given this, our work focuses on improving synchronous pipeline schedules as they ensure consistent semantics across different parallelism strategies.


The original GPipe~\citep{gpipe} simply divides a batch into several micro-batches, and its scheduling process has only two phases: the forward phase and the backward phase. The backward passes are executed only after the forward passes of all micro-batches within a batch are completed.
During the forward phase, the intermediate states of each micro-batch are enqueued into a FIFO queue $Q$. During the backward phase, these intermediate states are dequeued for their corresponding backward passes. Since the backward phase happens after all intermediate states are queued, GPipe exhibits an $O(M)$ memory consumption, where $M$ represents the number of micro-batches. 
TeraPipe~\citep{li2021terapipe} relies on the observation of causal language modeling, where the computation of a given input token only depends on its previous tokens, divides GPipe's micro-batch into multiple token spans, and replaces the FIFO queue with a last-in-first-out (LIFO) queue to ensure the correct computation of gradients in backward.
By using finer schedulable units (token spans), TeraPipe reduces the bubble ratio while being more memory-efficient than GPipe. Chimera~\citep{li2021chimera} adopts bidirectional schedule, where each device is responsible for processing multiple stages. While Chimera reduces the bubble ratio, each device has to store redundant parameters (as stages are not evenly distributed across devices), leading to increased memory usage.

Different from GPipe, 1F1B~\citep{megatron2021,fan2021dapple} alternates between forward and backward passes (adopting a 1F1B pattern) to keep the number of intermediate states in the FIFO queue $Q$ constant. Regardless of the number of micro-batches, 1F1B mitigates excessive memory usage. Based on 1F1B, 1F1B with interleaved stages (1F1B-I)~\citep{megatron2021} enlarges the number of pipeline stages and assigns each device multiple stages. By interleaving stages among devices, 1F1B-I reduces the bubble ratio at the cost of adding more communication operators and slightly increasing memory consumption. Zero-bubble-pipeline (ZB1P)~\citep{qi2024zero-bubble} divides the backward passes into obtaining weight and input gradients separately, which can achieve higher pipeline efficiency by delaying weight gradient computation and using dynamic programming to optimize the schedule. ZB1P nearly achieves zero-bubble pipeline efficiency but brings more memory footprint caused by delaying memory release. 1F1B methods are the most popular for training LLMs, yet still suffer from difficulties in balancing bubble ratio and memory footprint, which is the issue we want to solve.




\begin{figure*}[t]
  \centering
\includegraphics[width=\textwidth]{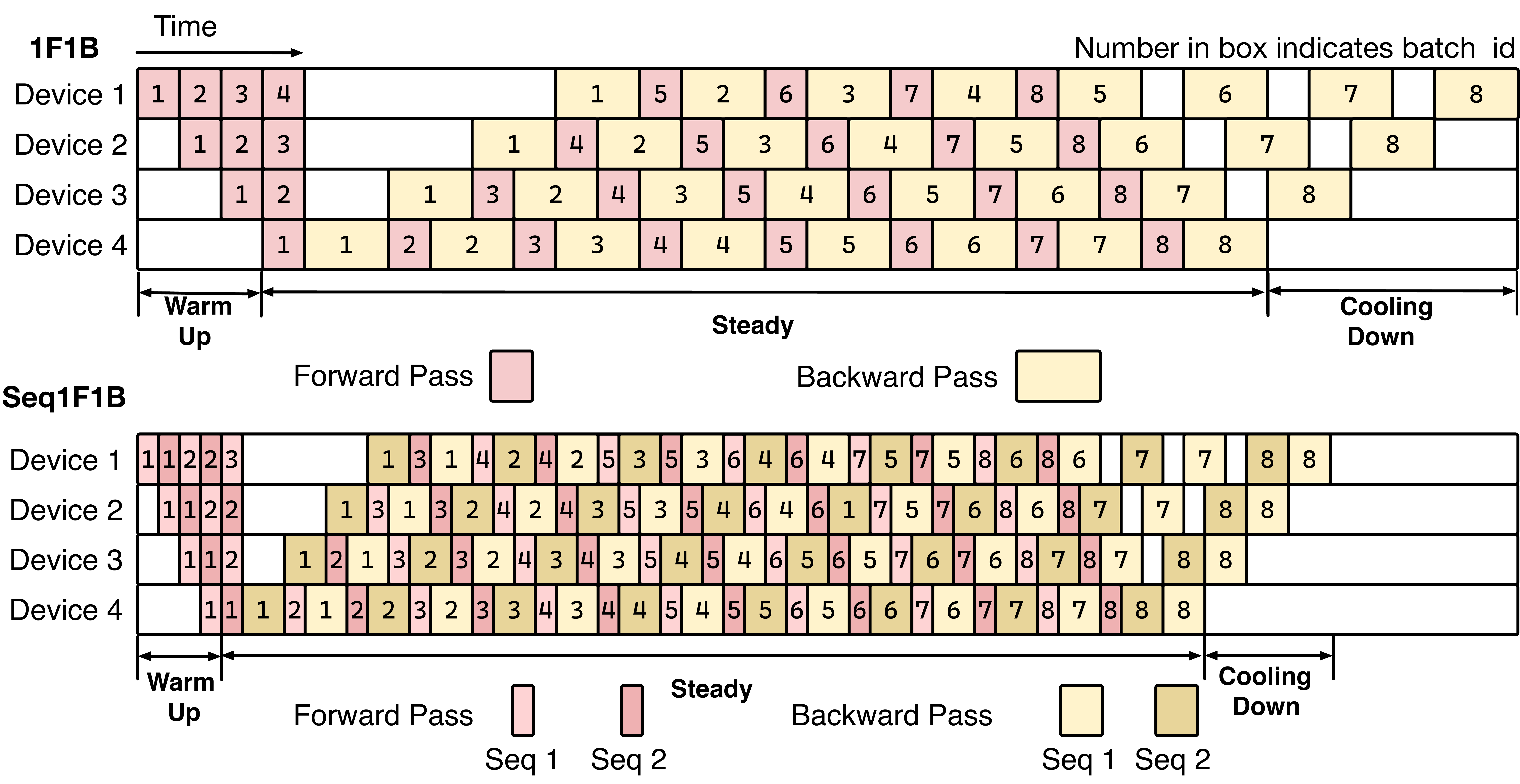} 
  \caption{Execution timeline for the 1F1B and Seq1F1B schedules. Blank spaces represent idle time, i.e. bubbles. The upper figure illustrates the original 1F1B schedule, where each micro-batch is labeled with an ID and bottom dashline represents the schedule phase of last device. The lower figure illustrates our Seq1F1B schedule, where the input is split into two sequences for better illustration. In Seq1F1B's illustration, light-colored areas represent the first sequence, while dark-colored areas represent the second sequence. Notice that the forward pass for the dark-colored sequence follows the light-colored sequence, whereas, for the backward pass, the dark-colored sequence precedes the light-colored sequence. }
  \label{fig:seq1f1b}
\end{figure*}

\section{Methodology}
\label{method}

In this section, we first give a preliminary overview to introduce the characteristics of the 1F1B schedule and language modeling. Then, we prove why it is feasible to schedule the pipeline of training LLMs at the sequence level for micro-batches in 1F1B. Finally, we explain how Seq1F1B works in detail and how it meets the exact semantics of original language modeling.


\subsection{Preliminary}
As shown in Figure~\ref{fig:seq1f1b}, \textbf{1F1B} includes three phases to train a batch of sequences: warm-up phase, steady phase, and cooling-down phase. 
Given $P$ devices (e.g., GPUs) to perform a 1F1B schedule to train $M$ micro-batches, with each device responsible for one pipeline stage, the size of PP is $P$.
After splitting the batch into $M$ micro-batches, during the warm-up phase, each device executes the forward passes of the first few micro-batches, and the number of forward passes $w_i$ executed by the $i$-th device is
\begin{equation}
\small
\centering
\text{w}_{i} = 
\begin{cases} 
P - i & \text{if } M > P \\
M & \text{if } M \leq P
\end{cases}
,\quad i \in [1, P],
\label{eq:warm-up-1f1b}
\end{equation}
When $M \leq P$, 1F1B degrades to the behavior of GPipe and does not process the steady phase. Otherwise, during the warm-up phase, a device responsible for an earlier stage performs one more forward pass than the device responsible for its subsequent stage. Each forward pass results in intermediate states enqueued in a FIFO queue $Q$ to be used later for the gradient computation of backward passes. 

During the steady phase, each device performs one forward pass and enqueues the resulting intermediate states into $Q$. After a device executes a forward pass, the device dequeues specific intermediate states from $Q$ and immediately executes a backward pass for gradient computation, where the ``1F1B'' name comes from. 
Note that, the bubble ratio is minimal during the steady phase, and the number of 1F1B passes in the steady phase is given by $M - \text{w}_{i}$. 
As $M$ increases, the proportion of the steady phase in the entire pipeline increases, which reduces the bubble ratio. 
After the steady phase, the 1F1B schedule enters the cooling-down phase, which is symmetric to the warm-up phase and involves executing the same number of backward passes as in the warm-up phase.

The primary optimization of 1F1B is to ensure that the memory consumption of intermediate states is independent of $M$. The peak memory consumption for intermediate states is determined by the number of items in the queue $Q$ at the end of the
warm-up phase, where each device holds $w_{i}$ intermediate states. Assuming the total
memory consumption of intermediate states is $A$, the peak memory consumption of the $i$-th device is $\frac{A\times w_{i}}{\sum_{j=1}^{P}w_j}$. During the steady and cooling-down phases,
this consumption does not increase since each backward pass frees the storage space for its associated intermediate states


\begin{figure*}[t]
  \centering
\includegraphics[width=\textwidth]{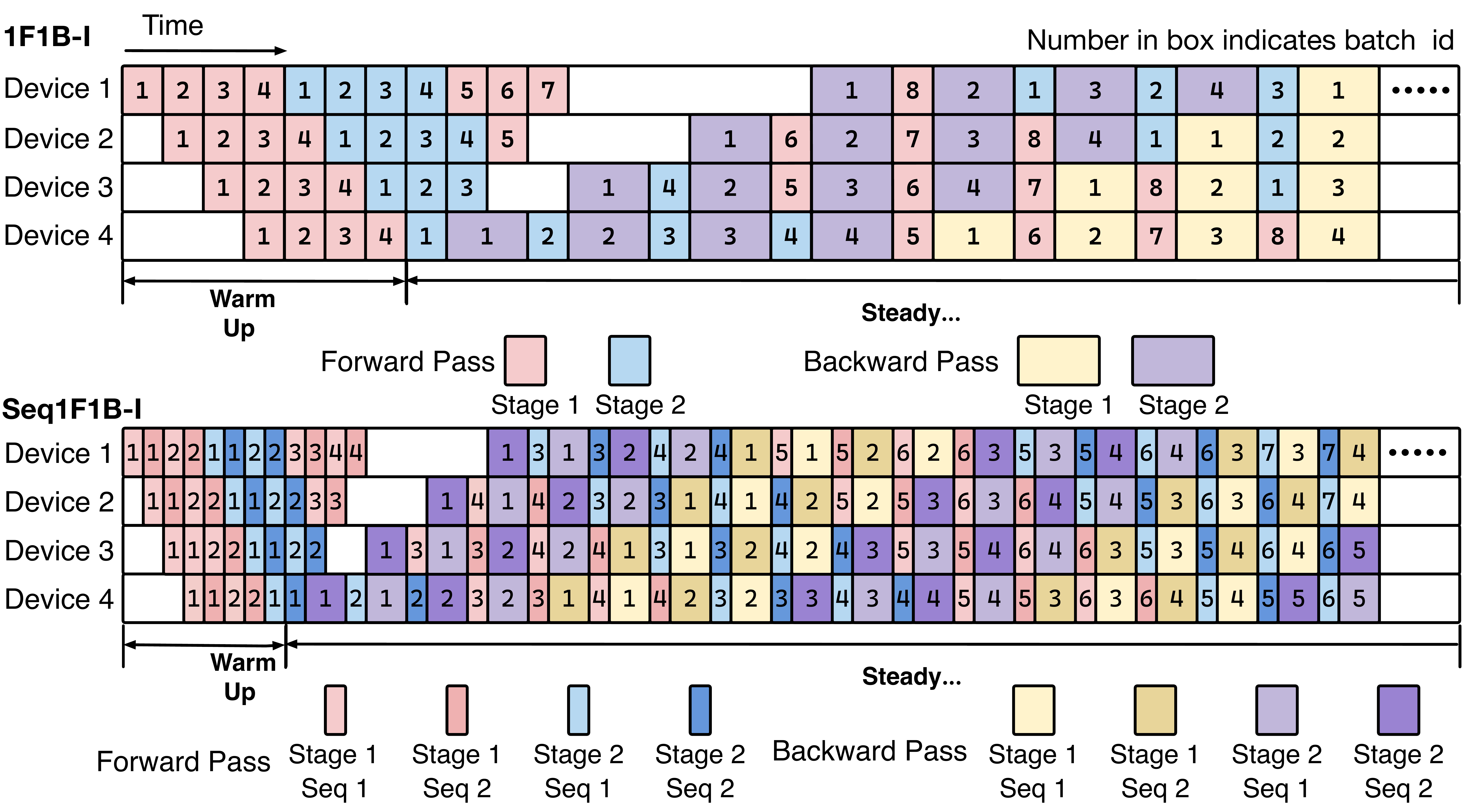}
\caption{Execution timeline for the 1F1B-I and Seq1F1B-I. The upper figure illustrates the 1F1B-I schedule, where each micro-batch is labeled with an ID, different colors distinguish the forward/backward passes of different stages. The lower figure shows the Seq1F1B-I schedule, where the input is split into two segments. In Seq1F1B-I, the light-colored areas represent the first sequence and the dark-colored areas represent the second sequence.} 
\label{fig:seq1f1b-I}
\end{figure*}

\textbf{Language modeling} is the most common unsupervised objective in training LLMs. 
In language modeling, each token is predicted sequentially while conditioned on the preceding tokens, embodying the principles of sequential generation, as formulated as 
\begin{equation}
\small
    P(\mathbf{x}) = \prod_{t=1}^{T} P(x_{t} \mid x_{1}, x_{2}, \ldots, x_{t-1}),
    \label{eq:lm}
\end{equation}
where $T$ is the sequence length. In the context of language modeling using Transformers, the causal attention mechanism ensures that each token in a sequence can only see its predecessors to process hidden states, including itself. 
Given a sequence of input token states $\{\mathbf{x}_1, \mathbf{x}_2, \ldots, \mathbf{x}_T \}$, the output of the attention mechanism for each token can be computed as follows. Each token states $\mathbf{x}_i$ is mapped into a query vector $\mathbf{q}_i$, a key vector $\mathbf{k}_i$, and a value vector $\mathbf{v}_i$, and output for each token $\mathbf{o}_i$ is computed by attending over all previous tokens as follows,
\begin{equation}
\small
    O_i = \text{softmax}\left( \frac{\mathbf{q}_i^{\top} \cdot [\mathbf{k}_1, \ldots, \mathbf{k}_i]}{\sqrt{d_k}} \right) [\mathbf{v}_0, \ldots, \mathbf{v}_i],
    \label{eq:softmax}
\end{equation}
where $d_k$ is the vector dimension.
Based on these characteristics, it is clear that to partition the Transformer computation across the sequence dimension must retain the key and value vectors of all preceding tokens.
The forward and backward passes also need to maintain a specific order. The forward pass of each token must follow the completion of its predecessor's computation, while the backward pass requires the subsequent token's gradients to complete its computation.
This computational dependency needs to be fully considered in sequence-level pipeline schedule.

\subsection{Framework of Seq1F1B}
From Figure~\ref{fig:seq1f1b}, we observe that the original 1F1B schedule cannot accommodate the splitting of micro-batches along the sequence dimension because the last stage needs to immediately execute a backward pass after forwarding a micro-batch. A straightforward adaptation method is to divide each original 1F1B micro-batch into $k$ segments and then execute a $k$F$b$B pipeline ~\citep{li2021terapipe}. Although this schedule can reduce some bubbles in 1F1B, it does not save memory usage.

To achieve a more efficient sequence-level 1F1B pipeline schedule, we propose Seq1F1B. 
Similar to 1F1B, the schedule of Seq1F1B is also divided into three phases: warm-up phase, steady phase, and cooling-down phase. 
During the warm-up phase, the number of sub-sequences of the $i$-th device is computed according to 
\begin{equation}
\small
\centering
\text{w}_{i} = 
\begin{cases} 
P - i - 1 + k & \text{if } M > P  \\
M & \text{if } M \leq P
\end{cases}
, \quad i\in [1, P],
\label{eq:warm-up-seq1f1b}
\end{equation}
where $P$ is the size of the PP and $k$ indicates 
the number of divisions of the sequence.
This equation ensures that the last stage can perform a backward pass on the last sub-sequence of the first micro-batch when entering the steady phase, and the device responsible for each stage performs one more forward pass than the device responsible for the subsequent stage.
Here, we construct a partially ordered queue $Q_{s}$, where each pop returns the tail sequence from the earliest enqueued intermediate states. This satisfies the FIFO principle in the batch dimension and the first-in-last-out (FILO) principle in the sequence dimension. In each step of the warm-up phase, devices execute one forward pass and enqueue the corresponding intermediate states of sub-sequences into $Q_{s}$. 
During the steady phase, after each device completes a forward pass, it dequeues intermediate states from $Q_{s}$ and performs a backward pass, following the standard 1F1B process, except that the units for forward and backward passes become a sub-sequence.
During the cooling-down phase, devices dequeue the remaining intermediate states from $Q_{s}$, perform backward passes for remaining subsequences and accumulate the gradient to ensure mathematical equivalent with other synchronous schedules.

From the timeline shown in Figure~\ref{fig:seq1f1b}, it is evident that the
Seq1F1B schedule offers a shorter execution time and significantly fewer bubbles,
compared to the original 1F1B schedule. Meanwhile, it can be seen that each device now has less memory consumption since the sub-sequence is smaller than the micro-batch. Another observation is that optimizations similar to ZB1P can also be applied to Seq1F1B by delaying the gradient computation associated with weights in the backward pass. For more details, we refer to our appendix.

\begin{table*}[t]
\small
    \centering
    \scalebox{0.95}{
    \begin{tabular}{rccccccc}
        \toprule
        Model & Number of & Attention  & Hidden  & Sequence  & PP & TP   & Number of \\
         Size & Layers & Heads & Size & Length & Size & Size & Micro-batches \\ 
        \midrule
        2.7B & 32 & 32 & 2560 & 16k / 24k / ~32k & 8 & 1 & 32 / 64 \\
        7B & 32 & 32 & 4096 & 32k / 64k / 128k & 4 & 8 & 16 / 32 \\
        13B & 40 & 40 & 5120 & 32k / 64k / 128k & 4 & 8 & 16 / 32 \\
        30B & 64 & 64 & 6144 & 32k / 48k / ~64k & 8 & 8 & 32 / 64 \\
        \bottomrule
    \end{tabular}}
    \caption{Settings used in experiments for training LLMs.}
    \label{tab:model-config}
\end{table*}

\subsection{Framework of Seq1F1B-I}
As shown in Figure~\ref{fig:seq1f1b-I}, 1F1B-I~\citep{megatron2021} achieves better efficiency by modifying the 1F1B schedule to support interleaved stages among devices. In 1F1B-I, each device is assigned multiple stages. Suppose we have $P$ devices and $V$ stages \(\{ s_{1}, s_{2}, \ldots, s_{V} \}\) in our pipeline, where $V$ is a multiple of $P$. The $i$-th device will handle $n$ stages \(\{ s_{i}, s_{i+P}, s_{i+2P}, \ldots, s_{i+(n-1)P} \}\), where \(n = \frac{V}{P}\). The number of warm-up micro-batches of each device $i$ in 1F1B-I is as follows,
\begin{equation}
\small
w_i =
(P - i ) \times 2 + (n-1) \times P, i\in [1,P],
\label{eq:warm-up-1F1B-I}
\end{equation}
After completing $P$ iterations of forward and backward passes, each device switches its context to the next stage that the device is responsible for. From Figure~\ref{fig:seq1f1b-I}, the above part shows a 1F1B-I pipeline with $P=4$ and $V=8$, in which each device handles 2 stages. The 1F1B-I schedule reduces the bubble ratio by interleaving stages among devices. However, this interleaving slightly increases memory consumption, as the number of warm-up micro-batches $w_{i}$ is greater than that of 1F1B.

Similar to 1F1B-I, Seq1F1B-I further modifies 1F1B-I to achieve a sequence-level schedule. From Figure~\ref{fig:seq1f1b-I}, Seq1F1B-I effectively reduces pipeline bubbles and the memory footprint of intermediate states compared to 1F1B-I. Seq1F1B-I defines the number of warm-up sub-sequences as
\begin{equation}
\small
w_{i} = (P - i ) \times 2 + (n - 1) \times P  + k - 1 , i\in [1,P],
    \label{eq:warm-up-seq1f1b-i}
\end{equation}
where $P$ is the size of the  PP and $k$ indicates 
the number of divisions of the sequence.
Using the partially ordered queue, Seq1F1B-I maintains a strict order of forward and backward passes as well as ensures the consistent semantics of gradient updates. 
From the perspective of pipeline bubbles, Seq1F1B-I outperforms both Seq1F1B and 1F1B-I. Besides, Seq1F1B-I requires slightly more memory than Seq1F1B but significantly less than 1F1B-I.

\subsection{Workload Balance}

In this section, we detail the strategy of sequence partition and workload balance consideration. Previous works, such as ~\citep{li2021terapipe}, have discussed strategies for sequence partitioning. To achieve an efficient pipeline schedule, the processing cost for each sub-sequence must be approximately equal to avoid pipeline bubbles. To this end, we design a computation-wise partition strategy by estimating the FLOPs of sequences and constructing a theoretical solution aiming to make the FLOPs of all sub-sequences as closely as possible.
For a input sequence $S = \{x_1, x_2,\cdots,x_n\}$, we devide it into $k$ segments $S=\{S_1, S_2, \cdots, S_k\}$.
Each segment has a length of $n_i$, where $\sum_{i=1}^k n_i = n$.
We expect the computational amount of each segment to be roughly the same, that is
\begin{equation}
\small
\begin{aligned}
\text{FLOPs}(S_1) & = \text{FLOPs}(S_2)  \\
& = \cdots = \text{FLOPs}(S_k)\\
& = {\text{FLOPs}(S)\over k}.
\vspace{-2em}
\label{eq:solver}
\end{aligned}
\end{equation}
Specifically, we use the method proposed in~\citep{hoffmann2022training} to estimate the FLOPs for each subsequence, formulated as
\begin{equation}
\small
\begin{aligned}
 \text{FLOPs}(S_i)&=2n_iP + 2Ln_i\left(\sum_{j=0}^{i} n_j\right)d, \forall i \in [1,k],\\ 
 \text{FLOPs}(S) &= 2nP + 2Ln^{2}d,
\end{aligned}
\label{eq:tflops}
\end{equation}
in which, $L$ is the number of layers, $d$ is the dimension of the model, and $P$ is the total number of parameters in the model. 
We have $k$ variables in Eq.~(\ref{eq:tflops}) and $k$ equations in Eq.~(\ref{eq:solver}), and thus we can set up the equation to get the optimal segmentation.




\begin{figure}
    \centering
    \includegraphics[width=\linewidth]{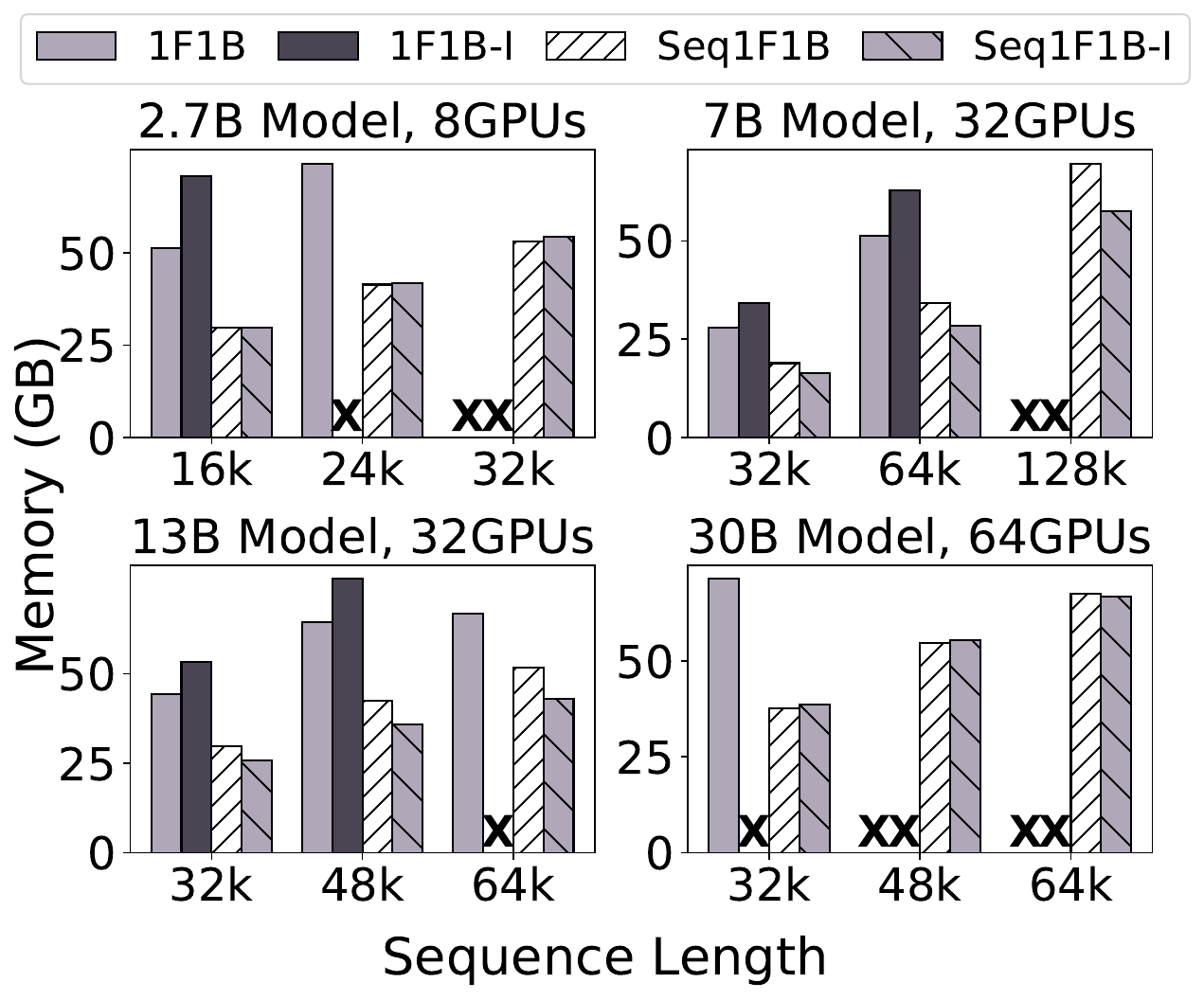}
    \caption{Peak Memory consumption of training a series of models
    under varying sequence lengths and fixed batch settings. ``X'' means experiments ran out of memory. We take the maximum memory consumption between all devices for better clarification.}
    \label{fig:exp-mem}
\end{figure}

\begin{table*}
\small
\centering
\scalebox{0.95}{
\begin{tabular}{llcccccc}
\toprule
\multicolumn{2}{c}{Model Size} & \multicolumn{6}{c}{2.7b} \\
\cmidrule{3-8}
\multicolumn{2}{c}{Sequence Length} & \multicolumn{2}{c}{16384}  & \multicolumn{2}{c}{24576}  & \multicolumn{2}{c}{32768} \\
 \cmidrule(lr){3-4}\cmidrule(lr){5-6}\cmidrule(lr){7-8}
 \multicolumn{2}{c}{Micro-batch} & 16 & 32 & 16 & 32 & 16 & 32 \\
\cmidrule(lr){1-2}\cmidrule(lr){3-4}\cmidrule(lr){5-6}\cmidrule(lr){7-8}
\multirow[c]{2}{*}{Throughput} & 1F1B & 32.0±0.0 & 37.1±0.0 & 27.0±0.0 & 31.4±0.0 & OOM & OOM \\
\multirow[c]{2}{*}{(Thousands} & 1F1B-I & 36.4±0.0 & \textbf{39.7±0.0} & OOM & OOM & OOM & OOM \\
 \multirow[c]{2}{*}{~Tokens/s)}  & Seq1F1B & \textbf{37.3±0.0} & 38.9±0.3 & \textbf{32.6±0.0} & \textbf{34.2±0.0} & \textbf{28.8±0.0} & \textbf{30.1±0.2} \\
 & Seq1F1B-I & \textbf{38.0±0.0} & 38.9±0.0 & \textbf{33.3±0.0} & \textbf{34.3±0.0} & \textbf{29.5±0.0} & \textbf{30.3±0.0} \\
\cmidrule(lr){1-2}\cmidrule(lr){3-4}\cmidrule(lr){5-6}\cmidrule(lr){7-8}
\multirow[c]{2}{*}{TFLOPS} & 1F1B & 96.9±0.0 & 112.3±0.0 & 95.5±0.1 & 111.1±0.1 & OOM & OOM \\
\multirow[c]{2}{*}{per device} & 1F1B-I & 110.3±0.1 & \textbf{120.2±0.1} & OOM & OOM & OOM & OOM \\
 & Seq1F1B & \textbf{113.1±0.0} & 117.8±0.8 & \textbf{115.2±0.1} & \textbf{120.9±0.1} & \textbf{116.5±0.1} & \textbf{122.0±1.0} \\
 & Seq1F1B-I & \textbf{115.2±0.0} & 118.0±0.0 & \textbf{118.0±0.1} & \textbf{121.3±0.1} & \textbf{119.4±0.0} & \textbf{122.7±0.0} \\
\bottomrule
\end{tabular}
}
\caption{2.7B GPT training experiments with PP size of 8 under 8$\times$ A100 setting.}
\label{tab:2_7b}
\end{table*}

\begin{table*}[t]
\small
\centering
\scalebox{0.95}{

\begin{tabular}{llcccccc}
\toprule
\multicolumn{2}{c}{Model Size} & \multicolumn{6}{c}{7b} \\
\cmidrule{3-8}
\multicolumn{2}{c}{Sequence Length} & \multicolumn{2}{c}{32768}  & \multicolumn{2}{c}{65536}  & \multicolumn{2}{c}{131072} \\
 \cmidrule(lr){3-4}\cmidrule(lr){5-6}\cmidrule(lr){7-8}
 \multicolumn{2}{c}{Micro-batch} & 8 & 16 & 8 & 16 & 8 & 16 \\
\cmidrule(lr){1-2}\cmidrule(lr){3-4}\cmidrule(lr){5-6}\cmidrule(lr){7-8}
\multirow[c]{2}{*}{Throughput} & 1F1B & 48.2±0.1 & 55.3±0.2 & 37.3±0.0 & 43.1±0.0 & OOM & OOM \\
  \multirow[c]{2}{*}{(Thousands}    & 1F1B-I & 53.0±0.3 & \textbf{56.3±0.4} & 41.7±0.1 & 44.7±0.0 & OOM & OOM \\
\multirow[c]{2}{*}{~Tokens/s)}  & Seq1F1B & \textbf{53.5±0.3} & 55.8±0.1 & \textbf{43.3±0.0} & \textbf{45.0±0.1} & \textbf{30.4±0.0} & \textbf{31.6±0.0} \\
 & Seq1F1B-I & 47.2±0.9 & 46.2±0.8 & 40.9±0.4 & 41.0±0.3 & 30.0±0.0 & 30.4±0.0 \\
\cmidrule(lr){1-2}\cmidrule(lr){3-4}\cmidrule(lr){5-6}\cmidrule(lr){7-8}
\multirow[c]{2}{*}{TFLOPS} & 1F1B & 99.7±0.2 & 114.5±0.4 & 107.5±0.0 & 124.0±0.1 & OOM & OOM \\
\multirow[c]{2}{*}{per device}  & 1F1B-I & 109.5±0.7 & \textbf{116.5±0.8} & 120.0±0.2 & 128.7±0.1 & OOM & OOM \\
 & Seq1F1B & \textbf{110.6±0.5} & 115.3±0.2 & \textbf{124.6±0.1} & \textbf{129.7±0.5} & \textbf{136.7±0.1} & \textbf{142.1±0.0} \\
 & Seq1F1B-I & 97.7±1.8 & 95.5±1.6 & 117.8±1.3 & 118.0±0.8 & 135.1±0.2 & 136.6±0.2 \\
\bottomrule
\end{tabular}
}
\caption{7B GPT training experiments with PP size of 4 and TP size of 8 under 32$\times$A100 setting.}

\label{tab:7b}
\end{table*}

\section{Experiments}
\label{exp}
\subsection{Experimental Settings}
\label{exp_setting}

In experiments, we measure Seq1F1B, Seq1F1B-I, 1F1B, and 1F1B-I under variable sequence lengths, different numbers of micro-batches, different numbers of GPUs, and different PP and TP sizes. Compared methods are as follows:

(1) Seq1F1B: Seq1F1B with computation-wise sequence partition strategy.

(2) Seq1F1B-I: Seq1F1B with interleaved stages and computation-wise sequence partition strategy.

(3) 1F1B/1F1B-I: 1F1B and 1F1B with interleaved stages in Megatron implementation.

(4) Seq1F1B w/o cwp: Seq1F1B without computation-wise sequence partition strategy.

(5) Seq1F1B-I w/o cwp: Seq1F1B-I without computation-wise sequence partition strategy.

All assessments are based on the
GPT model and model configurations are listed in Table~\ref{tab:model-config}.
All experiments focus on long-sequence training since a lot of work has mentioned its importance.
For hyperparameter configurations, we set the number of sequence splits to four and each device manages two stages in interleaved settings. Our implementation is based on the open-source Megatron-LM project~\citep{megatron2021} and ensures reproducibility. We adopt Megatron-V3~\citep{megatron-v3}'s tensor parallelism in all experiments since it is necessary for long sequence training.

Our experiments include three cluster settings: 
1) 1 node with 8 NVIDIA A100 SXM 80G GPUs interconnected by NvLink.
2) 4 nodes interconnected by a RoCE RDMA network and each node has 8 NVIDIA A100 SXM 80G GPUs interconnected by NvLink.
3) 8 nodes interconnected by a RoCE RDMA network and each node has 8 NVIDIA A100 SXM 80G GPUs interconnected by NvLink.
Each measurement in the experiment is repeated 100 times, and the standard deviation is recorded. 

\subsection{Main Results}

\begin{table*}[t]
\small
\centering
\scalebox{0.95}{
\begin{tabular}{llcccccc}
\toprule
\multicolumn{2}{c}{Model Size} & \multicolumn{6}{c}{13b} \\
\cmidrule{3-8}
\multicolumn{2}{c}{Sequence Length} & \multicolumn{2}{c}{32768}  & \multicolumn{2}{c}{49152}  & \multicolumn{2}{c}{65536} \\
 \cmidrule(lr){3-4}\cmidrule(lr){5-6}\cmidrule(lr){7-8}
 \multicolumn{2}{c}{Micro-batch} & 8 & 16 & 8 & 16 & 8 & 16 \\
\cmidrule(lr){1-2}\cmidrule(lr){3-4}\cmidrule(lr){5-6}\cmidrule(lr){7-8}
\multirow[c]{2}{*}{Throughput} & 1F1B & 28.9±0.1 & 33.4±0.1 & 25.3±0.1 & 29.3±0.1 & 22.6±0.1 & 30.0±0.0 \\
 \multirow[c]{2}{*}{(Thousands} & 1F1B-I & 32.2±0.2 & \textbf{34.4±0.1} & 28.2±0.2 & 30.6±0.1 & OOM & OOM \\
    \multirow[c]{2}{*}{~Tokens/s)}    & Seq1F1B & \textbf{32.9±0.1} & 34.3±0.1 & \textbf{29.5±0.1} & \textbf{30.8±0.0} & \textbf{26.7±0.0} & \textbf{27.8±0.0} \\
 & Seq1F1B-I & 29.7±0.4 & 29.8±0.3 & 28.0±0.2 & 28.3±0.1 & 26.4±0.1 & 26.8±0.1 \\
\cmidrule(lr){1-2}\cmidrule(lr){3-4}\cmidrule(lr){5-6}\cmidrule(lr){7-8}
\multirow[c]{2}{*}{TFLOPS} & 1F1B & 106.7±0.2 & 123.0±0.5 & 109.5±0.5 & 126.2±0.6 & 111.9±0.5 & 135.1±0.2 \\
 \multirow[c]{2}{*}{per device} & 1F1B-I & 118.6±0.6 & \textbf{126.9±0.4} & 121.9±0.7 & 132.2±0.4 & OOM & OOM \\
 & Seq1F1B & \textbf{121.2±0.2} & 126.6±0.3 & \textbf{127.3±0.4} & \textbf{133.1±0.2} & \textbf{132.5±0.0} & \textbf{137.9±0.0} \\
 & Seq1F1B-I & 109.7±1.4 & 110.0±1.1 & 121.0±1.1 & 122.1±0.4 & 130.6±0.3 & 132.8±0.3 \\
\bottomrule
\end{tabular}
}
\caption{13B GPT training experiments with PP size of 4 and TP size of 8 under 32$\times$ A100 setting.}
\label{tab:13b}
\end{table*}

\begin{table*}[t]
\small
\centering
\scalebox{0.95}{
\begin{tabular}{llcccccc}
\toprule
\multicolumn{2}{c}{Model Size} & \multicolumn{6}{c}{30b} \\
\cmidrule{3-8}
\multicolumn{2}{c}{Sequence Length} & \multicolumn{2}{c}{32768}  & \multicolumn{2}{c}{49152}  & \multicolumn{2}{c}{65536} \\
 \cmidrule(lr){3-4}\cmidrule(lr){5-6}\cmidrule(lr){7-8}
 \multicolumn{2}{c}{Micro-batch} & 8 & 16 & 8 & 16 & 8 & 16 \\
\cmidrule(lr){1-2}\cmidrule(lr){3-4}\cmidrule(lr){5-6}\cmidrule(lr){7-8}
\multirow[c]{2}{*}{Throughput} & 1F1B & 26.4±0.1 & 31.2±0.2 & OOM & OOM & OOM & OOM \\
\multirow[c]{2}{*}{(Thousands} & 1F1B-I & OOM & OOM & OOM & OOM & OOM & OOM \\
 \multirow[c]{2}{*}{~Tokens/s)}   & Seq1F1B & \textbf{31.3±0.1} & \textbf{33.1±0.2} & \textbf{28.2±0.1} & \textbf{29.6±0.1} & \textbf{25.5±0.0} & \textbf{26.8±0.0} \\
 & Seq1F1B-I & 28.0±0.4 & 28.4±0.2 & 26.5±0.2 & 27.1±0.2 & 24.8±0.1 & 25.2±0.1 \\
\cmidrule(lr){1-2}\cmidrule(lr){3-4}\cmidrule(lr){5-6}\cmidrule(lr){7-8}
\multirow[c]{2}{*}{TFLOPS} & 1F1B & 104.8±0.3 & 123.9±0.7 & OOM & OOM & OOM & OOM \\
 \multirow[c]{2}{*}{per device} & 1F1B-I & OOM & OOM & OOM & OOM & OOM & OOM \\
 & Seq1F1B & \textbf{124.5±0.2} & \textbf{131.5±0.6} & \textbf{129.4±0.3} & \textbf{135.6±0.3} & \textbf{132.6±0.0} & \textbf{139.2±0.0} \\
 & Seq1F1B-I & 111.1±1.6 & 113.0±1.0 & 121.5±1.1 & 124.2±0.8 & 128.6±0.3 & 130.9±0.6 \\
\bottomrule
\end{tabular}
}
\caption{30B GPT training experiments with PP size of 8 and TP size of 8 under 64$\times$ A100 setting.}
\label{tab:30b}
\end{table*}

In Figure~\ref{fig:exp-mem}, we compared the memory consumption of our method with that of 1F1B and 1F1B-I. As can be seen, our method consistently requires less memory across all settings. Notably, it can support training a 30B model on a 64$\times$ A100 cluster, which is impossible for the traditional combination of PP and TP. Additionally, we recorded TFLOPS (teraFLOPS) per GPU in our experiments to measure the hardware utilization of different methods. From Table~\ref{tab:2_7b}, \ref{tab:7b}, \ref{tab:13b} and \ref{tab:30b}, our method Seq1F1B outperforms 1F1B and 1F1B-I under almost all settings in both training throughput and teraFLOPS. 

However, as observed in Table~\ref{tab:7b}, \ref{tab:13b}, and \ref{tab:30b}, the Seq1F1B-I may have a performance degradation under multi-node settings. This could be due to the overly fine-grained interleaving of stage partitioning and input sequence partitioning, which also implies that more communication calls in TP (although the total communication volume remains unchanged), potentially leads to a decrease in performance.
Another observation is that the efficiency of Seq1F1B becomes more pronounced as the sequence length increases. This is because the computation time for each micro-sequence extends with longer sequences, thereby enhancing the benefits derived from sequence partitioning.

\subsection{Ablation Results}


We also conducted all experiments using Seq1F1B without computation-wise partitioning (Seq1F1B w/o cwp) and Seq1F1B-I without computation-wise partitioning (Seq1F1B-I w/o cwp) to evaluate the effectiveness of our computation-wise partition strategy. Under identical settings, employing the computation-wise partition strategy leads to performance enhancements ranging from approximately 10-30\% for Seq1F1B compared to simply splitting the sequence. 

Across all experimental scales, Seq1F1B consistently surpassed Seq1F1B w/o cwp in performance. Table~\ref{tab:ab_exp} highlights the ablation performance for a 2.7B model with a sequence length of 32k, demonstrating a performance boost of approximately 28\% due to the computation-wise partitioning.

\section{Conclusion}

\begin{table}[t]
\small
\centering
\scalebox{0.95}{
\begin{tabular}{lcc}
\toprule
          Method        & TFLOPS/device               & SpeedUp \\
\midrule
          Seq1F1B w/o cwp   & 94.8±0.1           & -     \\ 
          Seq1F1B           & 122.0±1.0 & \textbf{1.28} $\times$   \\
\midrule
          Seq1F1B-I w/o cwp & 103.5±0.1          & -     \\
          Seq1F1B-I         & 122.7±0.0 & \textbf{1.18}$\times$  \\
\bottomrule
\end{tabular}

}
\caption{The Ablation experiments are based on 2.7B GPT of sequence partitioning strategies, where ``w/o cwp'' indicates the absence of a computation-wise partitioning strategy.
}
\label{tab:ab_exp}
\end{table}
\

In this paper, we present Seq1F1B, an efficient 1F1B pipeline parallel schedule orienting to training Transformer-based LLMs on long sequences by decomposing the batch-level schedulable units used by typical 1F1B methods into more fine-grained sequence-level units. 
To achieve a better workload balance of the sequence-level pipeline, we design a computation-wise sequence partition strategy to partition the sequences well. Meanwhile, Seq1F1B can integrate with other pipeline parallel methods such as 1F1B with interleaved stage or zero-bubble-pipeline. Our evaluations demonstrate that Seq1F1B outperforms the 1F1B and 1F1B-I schedules regarding memory efficiency and training throughput under variable sequence lengths and model sizes. Moreover, Seq1F1B can support the efficient training of a 30B GPT model on sequences up to 64k in length using 64$\times$ A100 GPUs, without recomputation strategies, which is unachievable with existing pipeline parallel methods.
In the future, we will thoroughly combine our method with other distributed methods to achieve better LLM training acceleration. In addition, we will systematically release our code to support the community in training LLMs to process longer sequences more efficiently.

\section*{Limitations}

The current implementation of Seq1F1B is optimized for long-context training in LLMs, which may result in performance degradation when dealing with short context such as 4k/8k. We recommend using Seq1F1B in environments with limited communication bandwidth, as the PP  incurs fewer communication costs compared to other parallel strategies. 

\bibliography{custom}

\begin{thebibliography}{20}
\providecommand{\natexlab}[1]{#1}

\bibitem[{Anil et~al.(2023)Anil, Dai, Firat, Johnson, Lepikhin, Passos,
  Shakeri, Taropa, Bailey, Chen et~al.}]{anil2023palm}
Rohan Anil, Andrew~M Dai, Orhan Firat, Melvin Johnson, Dmitry Lepikhin,
  Alexandre Passos, Siamak Shakeri, Emanuel Taropa, Paige Bailey, Zhifeng Chen,
  et~al. 2023.
\newblock Palm 2 technical report.
\newblock \emph{arXiv preprint arXiv:2305.10403}.

\bibitem[{Buckman and Gelada()}]{long}
Jacob Buckman and Carles Gelada.
\newblock Compute-optimal {Context} {Size}.

\bibitem[{Fan et~al.(2021)Fan, Rong, Meng, Cao, Wang, Zheng, Wu, Long, Yang,
  Xia et~al.}]{fan2021dapple}
Shiqing Fan, Yi~Rong, Chen Meng, Zongyan Cao, Siyu Wang, Zhen Zheng, Chuan Wu,
  Guoping Long, Jun Yang, Lixue Xia, et~al. 2021.
\newblock Dapple: A pipelined data parallel approach for training large models.
\newblock In \emph{Proceedings of the 26th ACM SIGPLAN Symposium on Principles
  and Practice of Parallel Programming}, pages 431--445.

\bibitem[{Harlap et~al.(2018)Harlap, Narayanan, Phanishayee, Seshadri, Devanur,
  Ganger, and Gibbons}]{pipedream}
Aaron Harlap, Deepak Narayanan, Amar Phanishayee, Vivek Seshadri, Nikhil
  Devanur, Greg Ganger, and Phil Gibbons. 2018.
\newblock Pipedream: Fast and efficient pipeline parallel dnn training.
\newblock \emph{arXiv preprint arXiv:1806.03377}.

\bibitem[{Hoffmann et~al.(2022)Hoffmann, Borgeaud, Mensch, Buchatskaya, Cai,
  Rutherford, Casas, Hendricks, Welbl, Clark et~al.}]{hoffmann2022training}
Jordan Hoffmann, Sebastian Borgeaud, Arthur Mensch, Elena Buchatskaya, Trevor
  Cai, Eliza Rutherford, Diego de~Las Casas, Lisa~Anne Hendricks, Johannes
  Welbl, Aidan Clark, et~al. 2022.
\newblock Training compute-optimal large language models.
\newblock \emph{arXiv preprint arXiv:2203.15556}.

\bibitem[{Huang et~al.(2019)Huang, Cheng, Bapna, Firat, Chen, Chen, Lee, Ngiam,
  Le, Wu et~al.}]{gpipe}
Yanping Huang, Youlong Cheng, Ankur Bapna, Orhan Firat, Dehao Chen, Mia Chen,
  HyoukJoong Lee, Jiquan Ngiam, Quoc~V Le, Yonghui Wu, et~al. 2019.
\newblock Gpipe: Efficient training of giant neural networks using pipeline
  parallelism.
\newblock \emph{Advances in neural information processing systems}, 32.

\bibitem[{Jiang et~al.(2024)Jiang, Sablayrolles, Roux, Mensch, Savary, Bamford,
  Chaplot, Casas, Hanna, Bressand et~al.}]{jiang2024mixtral}
Albert~Q Jiang, Alexandre Sablayrolles, Antoine Roux, Arthur Mensch, Blanche
  Savary, Chris Bamford, Devendra~Singh Chaplot, Diego de~las Casas, Emma~Bou
  Hanna, Florian Bressand, et~al. 2024.
\newblock Mixtral of experts.
\newblock \emph{arXiv preprint arXiv:2401.04088}.

\bibitem[{Korthikanti et~al.(2023)Korthikanti, Casper, Lym, McAfee, Andersch,
  Shoeybi, and Catanzaro}]{megatron-v3}
Vijay~Anand Korthikanti, Jared Casper, Sangkug Lym, Lawrence McAfee, Michael
  Andersch, Mohammad Shoeybi, and Bryan Catanzaro. 2023.
\newblock Reducing activation recomputation in large transformer models.
\newblock \emph{Proceedings of Machine Learning and Systems}, 5.

\bibitem[{Li et~al.(2020)Li, Zhao, Varma, Salpekar, Noordhuis, Li, Paszke,
  Smith, Vaughan, Damania et~al.}]{li2020pytorch}
Shen Li, Yanli Zhao, Rohan Varma, Omkar Salpekar, Pieter Noordhuis, Teng Li,
  Adam Paszke, Jeff Smith, Brian Vaughan, Pritam Damania, et~al. 2020.
\newblock Pytorch distributed: experiences on accelerating data parallel
  training.
\newblock \emph{Proceedings of the VLDB Endowment}, 13(12):3005--3018.

\bibitem[{Li and Hoefler(2021)}]{li2021chimera}
Shigang Li and Torsten Hoefler. 2021.
\newblock Chimera: efficiently training large-scale neural networks with
  bidirectional pipelines.
\newblock In \emph{Proceedings of the International Conference for High
  Performance Computing, Networking, Storage and Analysis}, pages 1--14.

\bibitem[{Li et~al.(2021)Li, Zhuang, Guo, Zhuo, Zhang, Song, and
  Stoica}]{li2021terapipe}
Zhuohan Li, Siyuan Zhuang, Shiyuan Guo, Danyang Zhuo, Hao Zhang, Dawn Song, and
  Ion Stoica. 2021.
\newblock Terapipe: Token-level pipeline parallelism for training large-scale
  language models.
\newblock In \emph{International Conference on Machine Learning}, pages
  6543--6552. PMLR.

\bibitem[{Narayanan et~al.(2021{\natexlab{a}})Narayanan, Phanishayee, Shi,
  Chen, and Zaharia}]{pipedream-flush}
Deepak Narayanan, Amar Phanishayee, Kaiyu Shi, Xie Chen, and Matei Zaharia.
  2021{\natexlab{a}}.
\newblock Memory-efficient pipeline-parallel dnn training.
\newblock In \emph{International Conference on Machine Learning}, pages
  7937--7947. PMLR.

\bibitem[{Narayanan et~al.(2021{\natexlab{b}})Narayanan, Shoeybi, Casper,
  LeGresley, Patwary, Korthikanti, Vainbrand, Kashinkunti, Bernauer, Catanzaro
  et~al.}]{megatron2021}
Deepak Narayanan, Mohammad Shoeybi, Jared Casper, Patrick LeGresley, Mostofa
  Patwary, Vijay Korthikanti, Dmitri Vainbrand, Prethvi Kashinkunti, Julie
  Bernauer, Bryan Catanzaro, et~al. 2021{\natexlab{b}}.
\newblock Efficient large-scale language model training on gpu clusters using
  megatron-lm.
\newblock In \emph{Proceedings of the International Conference for High
  Performance Computing, Networking, Storage and Analysis}, pages 1--15.

\bibitem[{Qi et~al.(2024)Qi, Wan, Huang, and Lin}]{qi2024zero-bubble}
Penghui Qi, Xinyi Wan, Guangxing Huang, and Min Lin. 2024.
\newblock Zero bubble (almost) pipeline parallelism.
\newblock In \emph{The Twelfth International Conference on Learning
  Representations}.

\bibitem[{Rasley et~al.(2020)Rasley, Rajbhandari, Ruwase, and
  He}]{rasley2020deepspeed}
Jeff Rasley, Samyam Rajbhandari, Olatunji Ruwase, and Yuxiong He. 2020.
\newblock Deepspeed: System optimizations enable training deep learning models
  with over 100 billion parameters.
\newblock In \emph{Proceedings of KDD}, pages 3505--3506.

\bibitem[{Reid et~al.(2024)Reid, Savinov, Teplyashin, Lepikhin, Lillicrap,
  Alayrac, Soricut, Lazaridou, Firat, Schrittwieser et~al.}]{reid2024gemini}
Machel Reid, Nikolay Savinov, Denis Teplyashin, Dmitry Lepikhin, Timothy
  Lillicrap, Jean-baptiste Alayrac, Radu Soricut, Angeliki Lazaridou, Orhan
  Firat, Julian Schrittwieser, et~al. 2024.
\newblock Gemini 1.5: Unlocking multimodal understanding across millions of
  tokens of context.
\newblock \emph{arXiv preprint arXiv:2403.05530}.

\bibitem[{Shoeybi et~al.(2019)Shoeybi, Patwary, Puri, LeGresley, Casper, and
  Catanzaro}]{shoeybi2019megatron}
Mohammad Shoeybi, Mostofa Patwary, Raul Puri, Patrick LeGresley, Jared Casper,
  and Bryan Catanzaro. 2019.
\newblock Megatron-lm: Training multi-billion parameter language models using
  model parallelism.
\newblock \emph{arXiv preprint arXiv:1909.08053}.

\bibitem[{Touvron et~al.(2023)Touvron, Martin, Stone, Albert, Almahairi,
  Babaei, Bashlykov, Batra, Bhargava, Bhosale et~al.}]{touvron2023llama}
Hugo Touvron, Louis Martin, Kevin Stone, Peter Albert, Amjad Almahairi, Yasmine
  Babaei, Nikolay Bashlykov, Soumya Batra, Prajjwal Bhargava, Shruti Bhosale,
  et~al. 2023.
\newblock Llama 2: Open foundation and fine-tuned chat models.
\newblock \emph{arXiv preprint arXiv:2307.09288}.

\bibitem[{Vaswani et~al.(2017)Vaswani, Shazeer, Parmar, Uszkoreit, Jones,
  Gomez, Kaiser, and Polosukhin}]{vaswani2017attention}
Ashish Vaswani, Noam Shazeer, Niki Parmar, Jakob Uszkoreit, Llion Jones,
  Aidan~N Gomez, {\L}ukasz Kaiser, and Illia Polosukhin. 2017.
\newblock Attention is all you need.
\newblock In \emph{Proceedings of NeurIPS}.

\bibitem[{Yang et~al.(2021)Yang, Zhang, Li, R{\'e}, Aberger, and
  De~Sa}]{yang2021pipemare}
Bowen Yang, Jian Zhang, Jonathan Li, Christopher R{\'e}, Christopher Aberger,
  and Christopher De~Sa. 2021.
\newblock Pipemare: Asynchronous pipeline parallel dnn training.
\newblock \emph{Proceedings of Machine Learning and Systems}, 3:269--296.

\end{thebibliography}

\clearpage

\onecolumn
\appendix

\begin{figure*}[t!]
\centering
 \includegraphics[width=\textwidth]{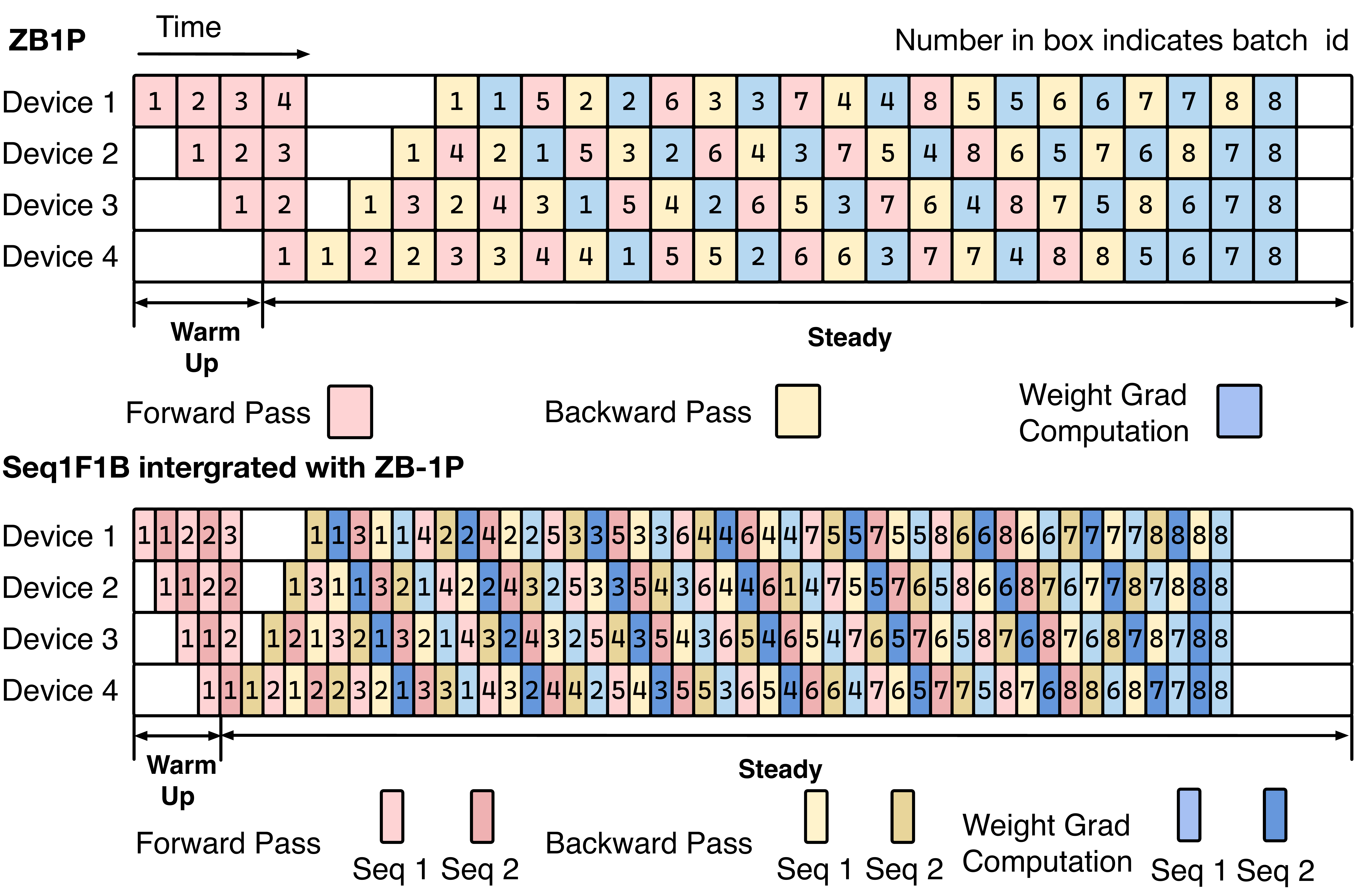}
\caption{Execution timeline for the zero-bubble-pipeline's ZB1P and Seq1F1B schedule intergrated with zero-bubble-pipeline's ZB1P. Each micro-batch is labeled with an ID and different colors to distinguish the forward/backward/weight computation of different stages. } 
\label{fig:seq1f1b-zerobubble}
\end{figure*}

\section{Appendix}
\subsection{Integration with Zero-bubble-pipeline}
\label{zero-bubble}

From Figure~\ref{fig:seq1f1b-zerobubble}, we can see that Seq1F1B can integrate with the ZB1P method and further reduce bubbles while reducing memory demands by splitting sequence. Such integration outperforms simple ZB1P in both memory demands and pipeline bubbles since sequence-level pipelines naturally have fewer bubbles. Furthermore, Seq1F1B can integrate with ZB2P and ZBV methods too. Theoretically, introducing a zero-bubble-pipeline to Seq1F1B should be more efficient. Even though, such a fine-grained handcraft schedule may have performance degradation under some settings. We hope our work inspires future work to solve this problem.

\end{document}